\documentclass[aps,prl,twocolumn,superscriptaddress,showpacs]{revtex4}

\usepackage{graphicx}
\usepackage{dcolumn} 
\usepackage{bm}      

\newcommand{\Tc}{{T$_c$~}}

\newcommand{\HRO}{{Ho$_2$Ru$_2$O$_7$~}}

\newcommand{\ub}{{$\mu_{B}$}}

\newcommand{\bfpsi}{\mbox{\boldmath$\psi$}}

\begin{document}

\preprint{APS/123-QED}

\title{Magnetic Ordering in the Spin-Ice Candidate {\HRO}}

\author{C.~R.~Wiebe}
\email{wiebecr@mcmaster.ca} \affiliation{Department of Physics and
Astronomy, McMaster University, Hamilton, Ontario L8S 4M1, Canada}
\affiliation{Department of Physics, Columbia University, New York,
New York 10027, USA}

\author{J.~S.~Gardner}
\affiliation{Department of Physics, Brookhaven National
Laboratory, Upton, New York, 11973-5000, USA} \affiliation {NIST
Center for Neutron Research, Gaithersburg, Maryland, 20899-5682,
USA}

\author{S.-J.~Kim}
\affiliation{Department of Physics and Astronomy, McMaster
University, Hamilton, Ontario L8S 4M1, Canada}

\author{G.~M.~Luke}
\affiliation{Department of Physics and Astronomy, McMaster
University, Hamilton, Ontario L8S 4M1, Canada}

\author{A.~S.~Wills}
\affiliation{Department of Chemistry, University College London,
20 Gordon Street, London, WC1H 0AJ, UK}

\author{B.~D.~Gaulin}
\affiliation{Department of Physics and Astronomy, McMaster
University, Hamilton, Ontario L8S 4M1, Canada}

\author{J.~E.~Greedan}
\affiliation{Department of Chemistry, McMaster University,
Hamilton, Ontario L8S 4M1, Canada}

\author{I.~Swainson}
\affiliation{NPMR, NRC, Chalk River, Ontario K0J 1J0, Canada}

\author{Y.~Qiu}
\affiliation{NIST Center for Neutron Research, Gaithersburg,
Maryland, 20899-5682, USA} \affiliation{Department of Materials
Science and Engineering, University of Maryland, College Park,
Maryland, 20742, USA}

\author{C.~Jones}
\affiliation{NIST Center for Neutron Research, Gaithersburg,
Maryland, 20899-5682, USA}

\date{\today}

\begin{abstract}

Neutron scattering measurements on the spin-ice candidate material
\HRO have revealed two magnetic transitions at T $\sim$ 95 K and T
$\sim$ 1.4 K to long-range ordered states involving the Ru and Ho
sublattices, respectively.  Between these transitions, the
Ho$^{3+}$ moments form short-ranged ordered spin clusters.  The
internal field provided by the ordered S=1 Ru$^{4+}$ moments
disrupts the fragile spin-ice state and drives the Ho$^{3+}$
moments to order.  We have directly measured a slight shift in the
Ho$^{3+}$ crystal field levels at 95 K from the Ru ordering.

\end{abstract}

\maketitle

Frustration, a condition which describes the inability of a system
to satisfy all of its individual interactions
simultaneously,\cite{Greedan} has become an important concept in
the realm of condensed matter physics, being applicable to a wide
range of phenomena such as high-\Tc superconductors, liquid
crystal phase transitions, and protein folding.  A renewed
interest in \emph{geometrically frustrated} magnets has resulted
from this general interest in frustration and the discovery of new
magnetic ground states.  One of these new states is the spin-ice,
which occurs on the pyrochlore lattice of corner sharing
tetrahedra with weak ferromagnetic coupling between rare-earth
ions subject to strong axial crystal fields.\cite{Gingras}  In
particular, the $\langle$111$\rangle$ anisotropy of these sites
promotes a ``two-in, two-out" low temperature spin arrangement
upon each tetrahedron, which is stabilized by dipolar
interactions.\cite{Hertog}  The resulting ground state has a
macroscopic entropy associated with the many ways that each
tetrahedron can satisfy this condition independently of the other
tetrahedra.\cite{Ramirez} The short-ranged order of the spins on
each tetrahedra maps onto the problem of proton ordering in water
ice.  Pauling first realized the significance of the specific heat
anomaly at the ice transition temperature as being due to the
disorder at each oxygen site.\cite{Pauling} An excellent agreement
has been found between the spin ice model and physical properties
including magnetization,\cite{Cornelius}, \cite{Petrenko} specific
heat,\cite{Ramirez} and neutron scattering experiments
\cite{Bramwell2} of the three spin ices, Dy$_2$Ti$_2$O$_7$,
Ho$_2$Ti$_2$O$_7$ and Ho$_2$Sn$_2$O$_7$.

Recently, a new spin-ice candidate has been discovered by Bensal
\emph{et al.} - Ho$_2$Ru$_2$O$_7$.\cite{Bensal}  Whereas other
spin ices of the formula A$_2$B$_2$O$_7$ only have one magnetic
species on the A site, in \HRO both A and B sites are magnetic:
Ho$^{3+}$ J = 8 spins and Ru$^{4+}$ S=1 spins. Previous studies on
the closely related pyrochlores in the series
R$_{2}$Ru$_{2}$O$_{7}$ (R = Y, Nd) have revealed that the
Ru$^{4+}$ moments order at higher temperatures (T $\sim$ 100 K).
\HRO shows an anomaly in the magnetic susceptibility which agrees
with these findings and suggests that the Ru$^{4+}$ moments order
at $\sim$ 95 K.\cite{Ito} However, this claim has not been
verified until this work. This letter details the study of \HRO by
neutron scattering to determine if the Ru$^{4+}$ moments order
and, if so, to investigate the effect of the internal field on the
Ho$^{3+}$ moments, which dominate the magnetic response.  We will
show that the Ru$^{4+}$ moments do order at $\sim$ 95 K into a
spin-ice like state of their own, while magnetic short range
correlations develop between the Ho$^{3+}$ moments as the
temperature is lowered further.  The internal field associated
with the Ru$^{4+}$ sublattice appears to be enough of a
perturbation upon the Ho$^{3+}$ ions to induce a low temperature
transition to a long-range ordered state at $\sim$ 1.4 K which is
not seen down to 50 mK of the other spin-ices.

We have made 20g of \HRO powder with less than 1$\%$ excess Ru
metal as determined by X-ray and neutron diffraction.  The
magnetic properties were measured between 2 K and 600 K using a
commercial SQUID magnetometer. Elastic neutron diffraction
measurements were performed with 2.37 and 2.0775 \AA~ neutrons at
Chalk River and NIST respectively from room temperature to 100 mK.
Inelastic neutron scattering was performed using various
wavelengths (3 to 9 \AA) at the Disk Chopper Spectrometer (DCS) at
NIST.\cite{Copley} Symmetry analysis calculations were carried out
using the program SARA{\it h}-Representational
Analysis.\cite{Wills-Sarah} Rietveld refinements were done using
Fullprof.\cite{Fullprof}

High temperature susceptibility data was fitted to the Curie Weiss
law and a Weiss temperature of -3(2) K was determined.  This is in
agreement with -4(0.5) K \cite{Bensal} found by Bensal \textit{et
al.} and is indicative of weak antiferromagnetic coupling.
Assuming that the response is largely due to holmium, the Curie
constant corresponds to an effective moment of 9.29(3) \ub, just
short of the expected value for the Ho$^{3+}$ $^{5}$I$_{8}$ ion of
10.6 \ub~ and again in agreement with Bensal \emph{et al.}

\begin{figure}[t]
 \linespread{1}
 \begin{center}
  \includegraphics[scale=0.32,angle=0]{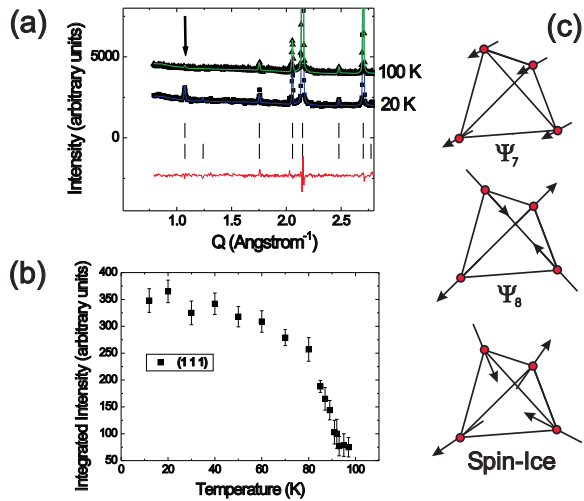}
  \caption{(a) Neutron scattering with 2.37 \AA~ neutrons at T = 100 K and T = 20
  K. The fits are to the crystal structure of
  Ho$_{2}$Ru$_{2}$O$_{7}$ (upper tick marks) and a magnetic structure (lower tick marks)
described in the text.  The residual of the 20 K fit is at the
bottom of the plot (R$_{p}$ = 1.99, R$_{wp}$ = 2.69, $\chi$$^{2}$
= 2.02 at 100 K; R$_{p}$ = 2.82, R$_{wp}$ = 3.73, R$_{mag}$ =
40.9, $\chi$$^{2}$ = 3.73 at 20 K). (b) The integrated intensity
of the magnetic (111) reflection (as indicated in figure 1(a)).
(c) $\bfpsi_7$ and $\bfpsi_8$ of the I.R. $\Gamma_9$, and the
spin-ice state which arises from equal proportions of these basis
vectors.}
  \label{Ru ordering}
 \end{center}
 \linespread{1.6}
\end{figure}

\begin{figure}[t]
 \linespread{1}
 \begin{center}
  \includegraphics[scale=0.32,angle=0]{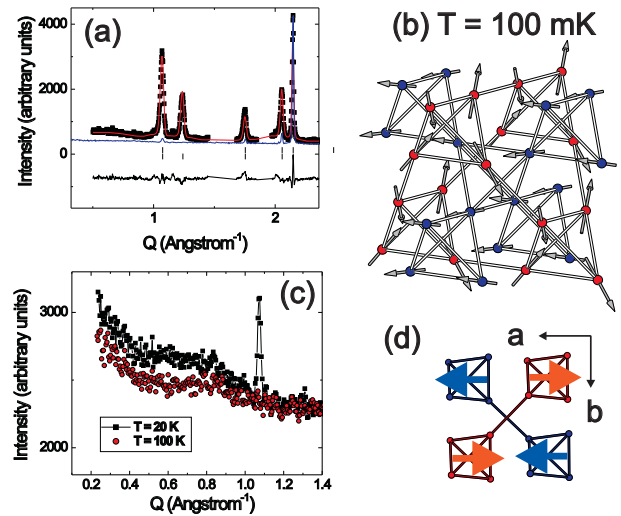}
  \caption{(a)  The diffraction data, fit (red line) and residual (black line) for 2.0775 \AA~ neutrons at 100 mK.  The upper tick marks are for the crystal structure and the lower for the magnetic (R$_{p}$ = 6.06, R$_{wp}$ = 7.85, R$_{mag}$ = 8.62, $\chi$$^{2}$ = 4.35).  The 20 K neutron data (blue line) is normalized to the (222) predominantly nuclear reflection.
  (b)  The magnetic structure of the Ru moments (blue) and Ho moments (red) at 100 mK.  For clarity, only a portion of the unit cell is shown, and the ordered moments are not drawn to scale.
  (c) Diffraction data with 2.37 \AA~ neutrons at T = 20 K and T = 100
  K, showing an intermediate state of SRO Ho$^{3+}$ above 1.4 K.  (d)  The magnetic structure at 100 mK can be thought of as a nearly collinear ferromagnet (Ru sites, in blue) and a spin-ice like state (Ho sites, in red) upon the different sublattices.  The net moment cancels from one sublattice to the next along one crystallographic direction.}
  \label{RuHoneutrons}
 \end{center}
 \linespread{1.6}
\end{figure}

Below 95 K, where a small field-cooled/zero-field-cooled
divergence in the susceptibility data is seen, magnetic Bragg
peaks appear which can be indexed with a ${\mathbf k}=0$
propagation vector.  These peaks are situated on top of diffuse
magnetic scattering at low \emph{Q}, which grows in intensity as
one cools (see figure 2).  This diffuse scattering is attributed
to regions of short-ranged magnetic order (SRO) from the Ho$^{3+}$
species.  Spin-ices have a characteristic diffuse scattering
profile which is indicative of the ferromagnetic SRO (ie. an
accumulation of scattering about \emph{Q} = 0 \cite{Kadowaki}).
However, an unambiguous determination of the nature of this
scattering requires further study, preferably on single crystals.
A slight broadening of the magnetic Bragg peaks with respect to
the nuclear, and the reduced ordered moment with respect to
expected S = 1 value (1.2(2) \ub~ as compared to 2 \ub) indicates
that not all of the Ru$^{4+}$ moments are ordered. From the (111)
magnetic peak, one can estimate the correlation length of the
Ru$^{4+}$ ordered spins to be $\sim$ 250 \AA, or about 25 unit
cells.  Further study of their ordering is not possible as the
diffuse scattering is dominated by that of the much larger
Ho$^{3+}$ moment.

\begin{figure}[t]
 \linespread{1}
 \begin{center}
  \includegraphics[scale=0.32,angle=0]{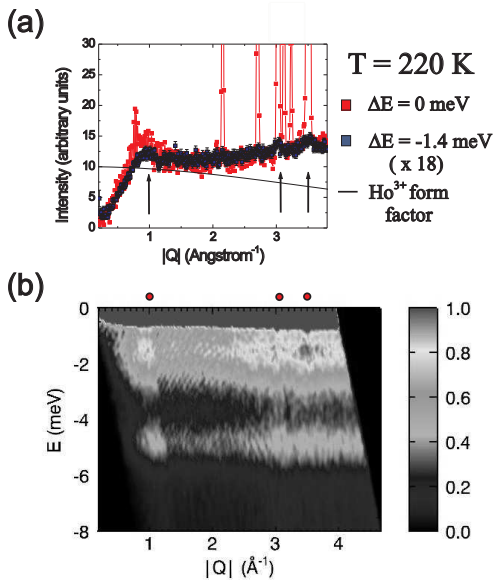}
  \caption{(a)  Integrated neutron scattering data at energy transfers of 0 meV and -1.4 meV and at 220 K (with a width of 0.6 meV) with the DCS and neutrons of
  $\lambda$ $\sim$ 3 $\AA$.  The positions of the (111), (422) and (440) reflections are noted,
as well as the Ho$^{3+}$ form factor.  (b)  Neutron scattering
contour plot at T = 220 K, showing dispersionless features at -1.4
meV and - 4.8 meV.\cite{dave}}
  \label{xtalfield1}
 \end{center}
 \linespread{1.6}
\end{figure}

\begin{figure}[t]
 \linespread{1}
 \begin{center}
  \includegraphics[scale=0.3,angle=0]{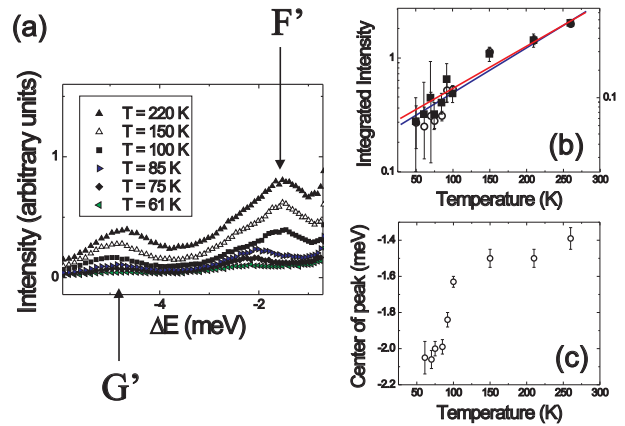}
  \caption{(a)  Integrated scans from $|$Q$|$ = 0.6 \AA$^{-1}$ to 4.0
\AA$^{-1}$ at 220 K.  Note the crystal field transitions,
indicated by F' and G'.
  (b)  Integrated intensities of the crystal field peaks, as fit to gaussians, as a function of temperature.  The open circles (blue fit) belong to the left scale (F' transitions) and the closed squares (red fit) belong to the right (G' transitions).
  (c)  Shift of the center of the gaussian peak for F' transitions.}
  \label{xtalfield2}
 \end{center}
 \linespread{1.6}
\end{figure}

The magnetic contribution to the powder neutron diffraction
spectrum of Ho$_2$Ru$_2$O$_7$ below 95 K can be well described by
ordering of the Ru$^{4+}$ moments according to the irreducible
representation $\Gamma_9$ of the space group
$Fd\bar{3}m$.\cite{symmetry note} This irreducible representation
has 6 associated basis vectors and may be thought of as involving
a ferromagnetic structure along the $a$-axis, $\bfpsi_7$, an
orthogonal antiferromagnetic structure, $\bfpsi_8$, and those
related by alternative choices of the lattice axis (see figure 1).
As neutron diffraction from a powder cannot distinguish the
orientation of these structures with respect to the cubic axes, we
restricted our analysis to the Hilbert space defined by $\bfpsi_7$
and $\bfpsi_8$. We note that equal proportions of $\bfpsi_7$ and
$\bfpsi_8$ correspond to a spin ice state with the propagation
vector ${\mathbf k}=0$. At 20 K the ordering is found to be
0.881$\bfpsi_7$ + 0.774$\bfpsi_8$ indicating that the Ru$^{4+}$
moments order with a spin ice-like local structure. It is
convenient to characterize these structures according to the
angle, $\theta$, that the moments make with the uncompensated
ferromagnetic component, in this case the $a$-axis. Thus a
collinear ferromagnet would have an angle of 0$^\circ$, the state
$\bfpsi_8$ an angle of 90$^\circ$, and spin-ice an angle of
109/2=54.5$^\circ$. A value of $\theta=41^\circ$ for the Ru
moments indicates that the structure is more ferromagnetically
collinear than the pure spin-ice state. The ordered component
(1.2(2)~$\mu_B$) is within error of the previous determinations of
1.36 \ub~ and 1.18 \ub~ for the ordered moments in
Y$_{2}$Ru$_{2}$O$_{7}$ and Nd$_{2}$Ru$_{2}$O$_{7}$ respectively,
but the proposed magnetic structures are different.\cite{Ito}  A
recent neutron scattering study of Er$_{2}$Ru$_{2}$O$_{7}$ has
revealed a planar structure involving Er$^{3+}$ and Ru$^{4+}$
ordered moments below 90 K, however we see no evidence for
Ho$^{3+}$ ordering at 90 K in \HRO.\cite{Taira}

Below 1.4 K, additional Bragg peaks appear in the diffraction data
(figure 2).  The data could only be well fitted by assuming that
both the Ru$^{4+}$ and Ho$^{3+}$ moments order according to the
representation $\Gamma_9$. Unlike the 20 K data, contributions
from both Ru and Ho sublattices were required, and the final
ordered moments are 1.8(6)~$\mu_B$ on the Ru and 6.3(2)~$\mu_B$ on
the Ho.  The Ru$^{4+}$ moments seem to be enhanced, but they are
still within error of the 20 K values. The refined moments were
oriented at 10$^\circ$ and 73$^\circ$ with respect to the
uncompensated component indicating that the Ho$^{3+}$ ordering
reduced the frustration of the Ru$^{4+}$ moments and increased
their collinearity. The Ho$^{3+}$ moments themselves are more
antiferromagnetic than ferromagnetic (with more $\Gamma_8$
character than $\Gamma_7$). Interestingly, the Ru$^{4+}$ moments
orient themselves such that they cancel one component of the
Ho$^{3+}$ spins, which in our definition is the a-axis. In the a-b
plane, one can think of the two sublattices as being
antiferromagnetically aligned, as shown in figure 2(d).

Bensal \textit{et al.} concluded that in \HRO, the long-range
dipolar interactions among the Ho$^{3+}$ spins do not destroy the
degeneracy of the spin-ice state, since the condition imposed by
den Hertog and Gingras,\cite{Hertog} J$_{eff}$ (effective nearest
neighbor energy scale) / D$_{NN}$ (dipolar energy scale) $\geq$
0.09 , is satisfied. However, we note that this is only true if
the Weiss constant is adjusted for Van Vleck paramagnetism and
demagnetization factors as found by Bramwell \textit{et al.} for
Ho$_2$Ti$_2$O$_7$.\cite{Bramwell}  More precise measurements are
needed on single crystals of \HRO to determine the Weiss constant
(and thus, J$_{eff}$). Our experiments clearly show that the
Ho$^{3+}$ moments do order, and do not form the spin-ice state.

High energy resolution inelastic neutron scattering was performed
on 20 g of sample on the DCS.  A representative spectrum at 220 K
is shown in figure 3(b). The dispersionless features at finite
energy transfer are a result of transitions between crystal field
levels at higher energies than those measured on the DCS (E $>$ 10
meV).  Figure 3(a) shows that these excitations do not follow the
form factor as predicted from isolated Ho$^{3+}$ moments, but they
are modulated in $\overrightarrow{Q}$ in a manner which follows
the diffuse scattering seen at $\Delta$E = 0 meV.  The scattering
tends to be peaked at values which correspond to the (111), (422)
and (440) reflections, which are ferromagnetic points within the
unit cell. This indicates that the Ho$^{3+}$ spins have net
ferromagnetic interactions, which is expected for a spin-ice, and
do not interact as isolated units. Future work is needed to
elucidate the Q-dependance of this data. It is unusual that such
correlations are seen at high temperatures, but not unprecedented,
as in the case of Tb$_{2}$Ti$_{2}$O$_{7}$, which has short-ranged
correlations up to at least 100 K.\cite{Gardner}

Using Ho$_{2}$Ti$_{2}$O$_{7}$ as a model and the notation adopted
by Rosenkranz \cite{Rosenkranz}, one can discuss the transitions
seen in figure 4. Transitions are observed at $\Delta$E = -1.5 meV
(denoted F') and -4.8 meV (denoted G'). Fitting these peaks to
gaussian functions, the temperature dependence can be plotted (see
figure 4(b)), and exponential behavior is noted with activation
energies of 105(4) K (F' transition) and 113(9) K (G' transition).
This corresponds to a separation of $\sim$ 10 meV from the ground
state.

\begin{figure}[t]
 \linespread{1}
 \begin{center}
  \includegraphics[scale=0.4,angle=0]{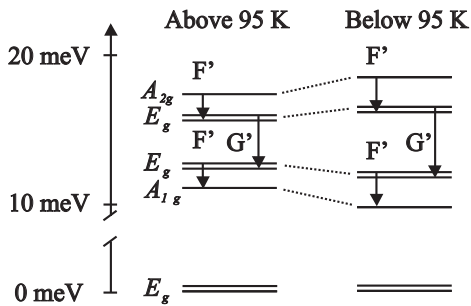}
  \caption{Schematic of the crystal field levels in \HRO (adapted from Rosenkranz \emph{et al.}\cite{Rosenkranz}).  The transitions F' and G' are noted.  As the Ru$^{4+}$ moments order, the crystal fields are
spilt slightly.
  The schematic is not to scale and the splitting is exaggerated for clarity.}
  \label{xtalfield3}
 \end{center}
 \linespread{1.6}
\end{figure}

There is a clear shift in the center of these peaks as a function
of temperature.  Figure 4(c) shows an increase in separation
between the levels below 95 K. The Ru$^{4+}$ ordering induces a
small internal field, which provides a Zeeman-like splitting of
the Ho$^{3+}$ levels as illustrated in figure 5.  Since there is a
small ferromagnetic component to the Ru$^{4+}$ ordering, this is
not a surprising result. It does show, however, that the Ru$^{4+}$
ordering has a measurable effect upon the Ho$^{3+}$ crystal
fields.  Although there is a short-ranged ordered state on the
Ho$^{3+}$ sites well above 1.4 K, it appears that the Ru$^{4+}$
ordering is enough of a perturbation of the fragile spin-ice state
to induce ordering upon the Ho$^{3+}$ site and drive the system to
order.  It has been suggested that the dominant interaction is
dipolar at low temperatures, with an interaction energy scale of
0.24 K.\cite{Bensal} This is reasonable to assume, given the
localized nature of the \textit{f} electrons of Ho$^{3+}$.
Superexchange pathways are likely to be complicated between the
two magnetic sublattices.

In conclusion, we find that \HRO is not a spin-ice, and has two
magnetic ordering transitions; with Ru$^{4+}$ and Ho$^{3+}$
ordering at $\sim$ 95 K and $\sim$ 1.4 K, respectively.  The
magnetic properties of the rare earth pyrochlores are the result
of a delicate balance between single ion anisotropy, exchange, and
dipolar coupling.\cite{Gingras} Although other rare earth
pyrochlores such as Er$_2$Ti$_2$O$_7$ \cite{Champion1} and
Gd$_2$Ti$_2$O$_7$ \cite{Champion2} order, it is found that the
corresponding structures vary considerably due to the roles played
by these complicated interactions.  We suggest that the Ho$^{3+}$
ordering found in \HRO is due to the small internal field produced
by the Ru$^{4+}$ ordering.  The subtle change in the Ho$^{3+}$
crystal field scheme that we have observed is convincing evidence
for this hypothesis.

\begin{acknowledgments}

C.~R.~Wiebe would like to acknowledge support from NSERC in the
form of a PDF.  The authors would like to thank the financial
support of NSERC, EMK, and CIAR.  This work utilized facilities
supported in part by the National Science Foundation under
Agreement No. DMR-0086210.  Work at Brookhaven is supported by the
Division of Material Sciences, U.~S.~Department of Energy under
contract DE-AC02-98CH10996.  The authors are also grateful for the
technical support of the NPMR staff at Chalk River, and Ross Erwin
at NIST.

\end{acknowledgments}


\begin{thebibliography}{99}
\bibitem{Greedan}J.~E.~Greedan, Chem. Mater. {\bf 10}, 3058
(1998).
\bibitem{Gingras}S.~T.~Bramwell {\em et al.}, Science {\bf
294}, 1495 (2001).
\bibitem{Hertog}B.~C.~den Hertog {\em et al.}, Phys. Rev. Lett. {\bf 84}, 3430 (2000).
\bibitem{Ramirez}A.~P.~Ramirez {\em et al.}, Nature {\bf 399}, 333
(1999).
\bibitem{Pauling}L.~Pauling, J. Am. Chem. Soc. {\bf 57}, 2680
(1935).
\bibitem{Cornelius}A.~L.~Cornelius {\em et al.}, Phys. Rev. B
{\bf 64}, 060406 (2001).
\bibitem{Petrenko}O.~A.~Petrenko {\em et al.}, Phys. Rev. B {\bf
68}, 012406 (2003).
\bibitem{Bramwell2}S.~T.~Bramwell {\em et al.}, Phys. Rev. Lett.
{\bf 87}, 047205 (2001).
\bibitem{Bensal}C.~Bensal {\em et al.}, Phys. Rev. B {\bf 66},
052406 (2002).
\bibitem{Ito}M.~Ito {\em et al.}, J. Phys. Chem. Solids {\bf 62},
337 (2001).
\bibitem{Copley}J.~R.~D.~Copley {\em et al.}, Chem. Phys. {\bf
292}, 477 (2003).
\bibitem{Wills-Sarah}
A.~S. Wills, Physica B {\bf 276}, 680 (2000), program available
from ftp://ftp.ill.fr/pub/dif/sarah/
\bibitem{Fullprof}
J. Rodr\'iguez-Carvajal, Physica B \textbf{192}, 55 (1993).
\bibitem{Kadowaki}H.~Kadowaki {\em et al.}, Phys. Rev. B {\bf 65},
144421 (2002).
\bibitem{dave}Data analysis was completed with DAVE, which can be
obtained at http://www.ncnr.nist.gov/dave/.
\bibitem{symmetry note}
We follow the labeling scheme used by Kovalev in
Ref.\onlinecite{Kovalev}.
\bibitem{Kovalev}
O. V. Kovalev, {\it Representations of the Crystallographic Space
Groups} Edition 2 (Gordon and Breach Science Publishers,
Switzerland, 1993).
\bibitem{Taira}N.~Taira {\em et al.}, J. Solid State Chem. {\bf
176}, 165 (2003).
\bibitem{Bramwell}S.~T.~Bramwell {\em et al.}, J. Phys. Cond.
Matt. {\bf 12}, 483 (2000).
\bibitem{Gardner}J.~S.~Gardner {\em et al.}, Phys. Rev. B {\bf
64}, 224416 (2001).
\bibitem{Rosenkranz}S.~Rosenkranz {\em et al.}, J. Appl. Phys.
{\bf 87}, 5914 (2000).
\bibitem{Champion1}J.~D.~M.~Champion {\em et al.}, Phys. Rev. B
{\bf 68}, 020401(R) (2003).
\bibitem{Champion2}J.~D.~M.~Champion {\em et al.}, Phys. Rev. B
{\bf 64}, 140407(R) (2001).



\end{thebibliography}
\end{document}